# Single Photon Emitters in Boron Nitride: More Than a Supplementary Material

M. Koperski[a,b*] K. Nogajewski[a], M. Potemski[a,b*]

[a]Laboratoire National des Champs Magnétiques Intenses, CNRS-UJF-UPS-INSA, Grenoble, France
[b]Institute of Experimental Physics, Faculty of Physics, University of Warsaw, Warsaw, Poland

*Corresponding authors at: Laboratoire National des Champs Magnétiques Intenses, 25, rue des Martyrs, B.P. 166, 38042 Grenoble Cedex, France (M. Potemski), School of Physics and Astronomy, University of Manchester, Oxford Road, Manchester M13 9PL, United Kingdom (M. Koperski)
*E-mail addresses*: Marek.Potemski@lncmi.cnrs.fr (M. Potemski), maciej.koperski@manchester.ac.uk (M. Koperski)

## ABSTRACT

We present comprehensive optical studies of recently discovered single photon sources in boron nitride, which appear in form of narrow lines emitting centres. Here, we aim to compactly characterise their basic optical properties, including the demonstration of several novel findings, in order to inspire discussion about their origin and utility. Initial inspection reveals the presence of narrow emission lines in boron nitride powder and exfoliated flakes of hexagonal boron nitride deposited on Si/SiO₂ substrates. Generally rather stable, the boron nitride emitters constitute a good quality visible light source. However, as briefly discussed, certain specimens reveal a peculiar type of blinking effects, which are likely related to existence of meta-stable electronic states. More advanced characterisation of representative stable emitting centres uncovers a strong dependence of the emission intensity on the energy and polarisation of excitation. On this basis, we speculate that rather strict excitation selectivity is an important factor determining the character of the emission spectra, which allows the observation of single and well-isolated emitters. Finally, we investigate the properties of the emitting centres in varying external conditions. Quite surprisingly, it is found that the application of a magnetic field introduces no change in the emission spectra of boron nitride emitters. Further analysis of the impact of temperature on the emission spectra and the features seen in second-order correlation functions is used to provide an assessment of the potential functionality of boron nitride emitters as single photon sources capable of room temperature operation.

## Introduction: boron nitride emitters among other single photon sources

Hexagonal boron nitride (hBN), a layered material known also as white graphite, is customarily used in cosmetic and metallurgic industry whereas its high purity single crystals promise the optical devices operating in the UV spectral range [1-11] and are extensively used in the research on atomically thin crystals [12-14], serving as nearly lattice matched and/or protecting layers for graphene and other two-dimensional systems [15-19]. The recent addition to explorations of hBN is the discovery of specific centres in this material, which rise a quantum light, the single photon emission (SPE) [20-27]. Known sources of quantum light are individual objects such as single atoms, molecules [28], defects in bulk crystals [29-31], nanocrystals, quantum dots [32, 33] and other localized/confined systems, such as, for example those lately reported to appear in thin layers of semiconducting transition metal dichalcogenides (s-TMDs) [34-38]. hBN is characterized by a large optical bandgap ($E_g$ = 5.5 eV) but most of SPE centres in this material can be effectively excited and emit light in the mid gap spectral range. The nature of SPE centres in hBN is unclear, though they are likely associated with the mid gap defects, similar to those found in other large band gap materials such as diamond [29, 30] or SiC [31]. The identification of SPE centres in hBN remains a challenge and has motivated the herein presented, extended characterization of those centres. Our studies are focused on SPE centres which are found in easily fabricated samples made of commercially available hBN powder (Sigma Aldrich supplier). Specific locations which give rise to sharp emission lines (SEL) and indeed display the SPE character can be relatively easily found in our structures with mapping of the micro-photoluminescence signal. These lines cover rather broad spectral range, located in visible region roughly between 500 and 700 nm. Single isolated lines, when investigated at multiple locations, reveal a pattern of occurrences, which can be described statistically in terms of a number of lines per energy unit. Such distribution overlaps well with spectra showing broad band-type emission in hBN powder samples, which may therefore arise from ensembles of individual emitters. Such interpretation of the origin of two different types of optical response seems to be further supported by the following inspection of the polarisation properties combined with measurements of excitation spectra. For the completeness of characterisation, we discuss later on the impact of magnetic field and temperature on SEL. The latter analysis is particularly useful for judging the quality of SPE, which is crucial for consideration of their potential functionalities.

**Appearance of single emitting centres in hBN powder and exfoliated hBN flakes**

The study presented here concern mostly the samples with hBN powder in form of grains of average diameter smaller than 150 nm deposited on Si/SiO$_2$ substrate. In this case, the process of sample preparation deserves some special attention, because, taking into account its simplicity, the resulting quality of the observed optical response is astonishing. After several try-outs (involving dissolving the powder in ethanol, annealing the samples in ambient conditions, etc.), a simple approach to distribute the hBN powder on a piece of a polydimethylsiloxane (PDMS) film, which acts as a transparent and flexible support typically used in the exfoliation technique of various layered materials (graphite, TMDs and also boron nitride), and to press it afterwards against the substrate for an extended period of time (about 30 minutes) turned out to be sufficient to observe bright SEL.

The samples have been characterised in a micro-optical setup, firstly by the photoluminescence (PL) mapping of their surface at helium temperature, about 4.2 K, with a laser beam focused to an area of about 1 μm diameter. Indeed, individual spots could be found, of the size below the spatial resolution of the set-up, which give rise to an optical signal in form of SEL. A unique pattern of lines represents each spot. Sometimes PL signal from such centres takes form of simple spectra with only a single line, but in most cases complex ensembles of lines are observed. They cover a relatively large spectral range in the visible region, reaching up to about 700 nm on the low energy side. As narrow lines can be found at the very edge of the transmission band of the filter used to suppress the laser light, clearly the observation of the high energy lines is limited by the energy of the laser excitation - usually Ar$^+$ 488 nm line. Particular SEL can be robust in intensity, but they require a rather high laser excitation power, probably due to sub-bandgap excitation leading to low absorption of the laser light. Typically, 1 − 2 mW of the laser power was delivered to the surface of sample in the mapping experiments, which is comparable to the excitation power typically used for quasi-resonant (below the band gap) excitation of, e. g., semiconductor quantum dots.

Representative results obtained from the μPL mapping of the hBN powder sample are presented in the left panel of **Figure 1**. The colour map illustrates the spatial distribution of the intensity monitored at a particular energy, 2.374 eV. Therefore, bright spots reveal places, where a spectral line appears at that chosen energy value. Three selected spectra originating from this map come from places marked as M1 - M3. They represent a distinct character of each emitting centre, including the case with only a single spectral line (M2). The distribution of the emitters on the surface of the sample is rather scarce - they tend to appear as single and isolated spots, as demonstrated by the colour map.

Even though the presence of the isolated emitting centres in the hBN powder samples is beyond doubts, one could still argue that their origin may not be related to hBN at all. A scenario that the chemical substances used at any step in the production of the hBN powder play a substantial role is sound and cannot be *a priori* excluded. A possible route to shed some light on this issue could be the examination of hexagonal hBN flakes obtained from bulk crystals - a material of well-established crystal structure and composition. Therefore, the hBN flakes exfoliated with a standard method developed mostly during studies of TMDs and graphene structures, have been deposited on Si/SiO$_2$ substrates to undergo similar mapping experiments as the hBN powder samples. In the right panel of **Figure 1**, the results of investigation of a selected bulk flake are summarised. The image from an optical microscope shows, based on the colour contrast with respect to the substrate, that the flake is thick and composed of multiple segments. Most of the characteristics of the flake's shape and structure can be reproduced in the mapping experiment by monitoring, for instance, the intensity of the scattered laser light at the consecutive spatial positions. Such faithful reproduction of the flake's geometry allows us to quite accurately establish a relation between spots in the map and particular places visible in the optical image with precision determined by the size of the laser spot. The centres emitting narrow spectral lines appear also in exfoliated flakes, as seen in the spectra originating from spots P1 - P3 indicated in the image and in the map. Generally, the spectra are overall similar to the ones obtained for hBN powder sample with the exception that the narrow lines overlap with a broad background of significant intensity, which constitutes a disadvantage for further optical studies. The origin of the background signal is most likely related to the bulk luminescence. Even though the bulk emission is expected to reside predominantly in the UV region, its tails may reach into visible spectral range. The observation of SEL in exfoliated hBN flakes provides a solid argument in favour of the hypothesis, that their origin is related to defects in hBN material. Yet, due to superior quality of the spectra obtained for hBN powder samples, they were used in all the proceeding investigations.

As a first step towards the characterization of the optical properties of the emitting centres in hBN, we examine the dependence of their emission spectra on the energy of the laser excitation. In **Figure 2** three spectra for the same location are presented, obtained under the excitation of 488 nm and 514 nm $Ar^+$ laser lines and 566 nm rhodamine dye laser line with a comparable excitation power of about 0.5 mW. An evident influence of the excitation energy on the intensity of individual lines can be observed. Multiple lines are seen in all three spectra at exactly the same energy, which provides a direct evidence that they appear indeed due to photoluminescence and not a Raman scattering process. Some of the lines clearly exhibit markings of a resonant character, for instance the most robust line in the spectrum for 566 nm excitation is barely visible in the spectrum for 488 nm excitation. Besides that, particular lines can demonstrate various dependence on the excitation energy. The intensity of the line **N˚ 1** is roughly constant for all three excitations. The line **N˚ 2** has an apparent maximum of intensity for the intermediate excitation. At the same time, the intensity of line **N˚ 3** progressively decreases when the excitation energy approaches the emission energy. Such diversified behaviour of the SEL must signify a complex energetic structure of the associated defects, which should be studied systematically and in details. Some preliminary attempts to unveil the energetic landscape and the nature of the light-matter coupling for the emitting centres in hBN are discussed here through the analysis of the excitation spectra combined with the inspection of polarisation properties.

From the point of view of more advanced spectroscopic experiments, as well as potential utility of the emitting centres in hBN, the matter of their stability plays a fundamentally important role. Also in this aspect, the hBN emitters provide a ground for further exploration. A vast majority of them give rise to narrow lines, which exhibit a perfect stability in temporal domain. That statement is valid for short sub-second timescale, in which no trace of telegraphic noise of the emission lines is observed, as well as longer periods of several hours, when no deterioration of the optical signal occurs during long-time experiments. A temporal evolution of a spectrum recorded for such a stable emitter is shown in left panel of **Figure 3**. A small number of emitting centres exhibit peculiar blinking effects, occurring in the timescale of seconds or even tens of seconds. In such a case, one can distinguish two states of the emitter, which clearly cannot coexist at the same time. An example of an emitter exhibiting this kind of blinking is presented in the right panel of **Figure 3**. Similar effects in other systems, such as semiconductor quantum dots, are usually related to the charge fluctuations, however they typically occur in orders of magnitude shorter timescales. The investigation of the switchable centres may allow the identification of charged states as well as grant an insight into the dynamics of local variation of the electric field in the vicinity of the emitting centres, which may prove a valuable asset in uncovering their characteristics through subtler spectroscopic tools, such as charge noise spectroscopy **[39]**.

To conclude the initial characterisation of the emitters in hBN it is noteworthy to mention, that SEL are not the only type of optical response observed in hBN powder samples. At multiple spots, also broad features appear at the energy covering similar region as the emission in form of sharp lines. In order to provide a more quantitative measure of comparison, the statistics of the appearance of the narrow lines at particular energy for about 80 emitting centres has been investigated. The results, in form of a histogram, are presented in **Figure 4**. On top of the statistical data, a spectrum is presented showing an example of a broad feature, one of the most commonly found in the hBN powder samples. The observed correspondence may serve as a motivation to consider a hypothesis, that apart from rare individual emitting centres also ensembles of them, which give rise to broad emission bands, exist in our samples.

**Characterisation of resonances by the photoluminescence excitation and polarisation resolved experiments**

Further information about the character of the emitting states can be found in the excitation spectra, which are presented in **Figure 6** together with their emission counterparts. Here, the intensity of individual lines is monitored, when the rhodamine dye laser energy is continuously tuned through the available spectral range, from 560 to 610 nm.

The PL spectrum for a selected location is composed mainly of five lines in the presented energy range and the PLE spectra are shown for each of them. An immediate observation is that the character of the excitation spectra may be qualitatively different for particular lines. The lower energy doublet, two PL lines between 1.75 and 1.8 eV, arises apparently due to non-resonant states, in the sense that the intensity of this pair of lines remains significant through the full excitation energy range. On top of that, both lines exhibit a certain type of resonant response. In the PLE spectrum of the lower energy line one can distinguish several rather weak and broad resonances while for the more robust, higher energy line an extensive band appears in the excitation spectrum. On the other hand, the triplet of lines, seen around 1.9 eV, reveals strictly resonant characteristics. For these three lines, narrow resonances appear with the linewidth comparable to the one of the PL lines, a feature most clearly pronounced for the lowest energy line of the triplet. Outside of the resonant energies, all three lines almost completely vanish from the PL spectrum, which enables to a large degree a selective and independent excitation of these emission lines.

In order to complete the characterization of the intrinsic properties of the emitting centres in hBN we analyse their polarisation properties measured for the same spot, which served as the example to present the PLE studies. We demonstrate the polarisation data in **Figure 6** separately for resonant and non-resonant emission lines. All five emission lines appearing in the PL spectrum are completely linearly polarised, each line along its own intrinsic axis. This signifies, that the linear polarisation of these lines originates from the internal properties of the emitting states, determined most likely by anisotropic structure of particular defects. At this point, there is no particular proof that all the lines seen in the spectrum originate from the same emitter, though it might as well be so. If that is the case, the linear polarisation data presented here may serve as an indication of the presence of fine structure splitting (splitting of states arising due to anisotropy of electronic states), as the two highest energy lines of the resonant triplet are polarized perpendicularly to each other and the splitting between them is rather small (3.2 meV).

SEL in hBN reveal also sensitivity to the direction of the linear polarisation of the excitation. Clearly, the absorption of light by the emitter can be tuned by rotating the linear polarisation axis of the laser, as seen in **Figure 6c**, in which the intensity of a selected resonant and non-resonant line is presented as a function of the linear polarisation angle for both excitation and detection as a colour plot. This finding, combined with the results of the PLE study, shows that there exist at least two possible ways of distinguishing individual emitters from ensembles, which exploit the excitation selectivity related to the laser energy and polarisation.

**Influence of external conditions: temperature and magnetic field dependence**

We proceed with the analysis of the optical properties of SEL with the discussion of the dependence of the emission spectra on varying external conditions such as temperature and magnetic field. In **Figure 7a** the evolution of the PL spectrum is presented in a temperature range from 10 to 296 K. The intensity, emission energy and linewidth is analysed for a selected line at 1.998 eV and the results are presented in **Figure 7b**. The influence of temperature on the properties of PL lines is rather standard for localized emitting centres in solids. The intensity is quenched, the emission energy is red-shifted and the lines are broadened at higher temperature. Apart from this typical evolution of individual narrow lines, an appearance of an additional feature becomes apparent, when the temperature of the sample exceeds 200 K. A broad band suddenly emerges covering completely the visible spectral region, where the narrow lines are found. The intensity of this background signal grows drastically with further increase of the temperature. The presence of a similar feature is commonly observed for various colour centres in wide gap materials. It is attributed to the phonon assisted transitions which accompany the zero-phonon line. However, this may not be the case for emitters in hBN powder. Firstly, the broad band does not appear at all at low temperatures, which would imply that phonons responsible for its emergence become involved at temperatures as high as 200 K, which is unlikely. Considering the fact that the intensity of the background emission grows in time, when exposing the emitter with laser excitation for extended periods of time at a set temperature, we currently consider the possibility that it comes from damaging the crystal due to heating it with the laser of a significant power.

The results of the magneto-optical studies are much more puzzling, because SEL in hBN powder (as well as exfoliated hBN crystals) appear to be completely insensitive to the magnetic field in the range up to 14 T. The results of measurements for a selected emitter in hBN powder sample are presented in **Figure 7c**. No signature of a field-induced splitting is observed for any of the lines nor a noticeable impact of the field on their intensity. This property is common for all measured emitting centres. Any possible explanations of this observation currently remain highly speculative. Nonetheless, we would like to point out two simple concepts leading to the disappearance of Zeeman effects for optical transitions; see **Figure 8**. The first scenario is based on a coupling between electrons. Let us imagine that the resonances seen in our experiments are related to optical transitions between vacuum and few electron states. If the involved electrons would form antiferromagnetically coupled pairs, then the energy of the electronic states could be magnetic field independent. Then, one can expect an appearance of higher energy triplet state, which in the simplest case of two electrons with 1/2 spin splits into three components. A feasible route to verify this hypothesis is extension of the magneto-optical studies to higher fields in search for, e. g. anti-crossings induced by a potential transformation of the ground electronic state from singlet into triplet configuration (or similar). An alternative idea simply takes into consideration a Zeeman splitting of individual electronic states of exactly the same magnitude, what leads to the compensation of the magnetic field splitting for the optical transitions. If the optical selection rules would favour the transitions exclusively between the two higher energy and the two lower energy components of the split electronic states, then the resonances seen in the optical spectra could be insensitive to the magnetic field.

**Photon correlation study - single photon emission from helium to room temperature**

Second-order correlation functions measured for single, well isolated PL lines from defect centres in hBN clearly confirm their single photon emission character. A robust anti-bunching is seen, which reveals its characteristic time to reside in tens of nanoseconds timescale. This value constitutes the upper limit for the PL decay time and is comparable with other single photon emitters in solids. Another feature, which is observed in the correlation functions in longer timescales (of the order of microseconds) is a bunching with unusually large amplitude. That is a peculiar trait of the SEL in hBN crystals of currently unknown origin. A bunching in a correlation function may appear due to charge fluctuations, which induce small variations in the energy of the electronic states relevant for the optical transitions. Alternatively, a bunching may also arise in multilevel systems, where the transitions between particular states are accounted for by rate equation models. We believe the latter explanation is more likely to be valid for SEL in hBN, based on the decent stability of their PL lines and the typical multi-line optical response. Photon correlation measurements were done in broad temperature range from helium up to room temperature. The experimental correlation functions, presented in **Figure 9**, were described by the following formula:

$$g_2(\tau) = 1 - A_1 exp\left(-\frac{|\tau|}{t_1}\right) + A_2 exp\left(-\frac{|\tau|}{t_2}\right)$$

where $t_1$ is an exponential temporal constant characterising the antibunching, $t_2$ is an exponential temporal constant characterising the long timescale bunching, $A_1$ is the antibunching depth and $A_2$ is the bunching amplitude. The temperature dependence of the parameters is presented in **Figure 9b**. The quality of SPE of SEL in hBN is exceptionally good. It may be represented by the renormalized value of the correlation function at zero-time delay: $g_2(0) \to \frac{g_2(0)}{1+A_2} = \frac{1-A_1+A_2}{1+A_2}$. The renormalisation is necessary to preserve the interpretation of the value of the $g_2(0)$ parameter as the probability of emitting more than one photon in the presence of a bunching of significant amplitude. The excellent quality of the SEL in hBN is seen in the low value of the renormalized $g_2(0)$, especially at low temperature (roughly below 200~K) when it does not exceed 5 %. At higher temperature, the value of the renormalized $g_2(0)$ increases significantly due to the emergence of the spectrally broad background signal. It is important to stress that this observation does not necessarily imply the deterioration of the SPE capabilities of SEL in hBN at higher temperature. It simply signifies that the appearing background signal does not come entirely from the same transition which at low temperature appears in form of a narrow, well isolated line. An example of such a situation could be a temperature activated emergence of large phonon sidebands, which would accompany each individual narrow line. Their merging into a singly extensive PL band could definitely limit the possibility of detecting signal exclusively from an individual optical transition. Nevertheless, the progress in understanding the origin and properties of SEL in hBN may point out ways to better isolate the luminescence coming from individual PL lines at higher temperatures and restore the almost perfect character of SPE even up to room temperature.

Another parameter which is evidently sensitive to the variation of the temperature is the amplitude of the bunching. It diminishes noticeably when increasing the temperature from 10 to 200 K, but given the uncertain origin of the bunching feature this observation is difficult to interpret. However, we would like to remark, that the rise of the temperature may substantially increase the ratio between the non-radiative and radiative processes. Therefore, the dynamics of the population and depopulation of the emitting states can be expected to be sensitive to temperature. The two remaining parameters, i. e. characteristic temporal constants related to the antibunching and the bunching appear to be practically independent on the temperature.

Finally, the photon correlation function was measured at room temperature for SEL in hBN at a new location to demonstrate the possibility of room temperature operation. The $g_2(\tau)$ function and the PL spectrum showing the correlated emission line is presented in **Figure 9c**. These data demonstrate that despite the high temperature background one can still observe SPE from SEL in hBN in ambient conditions. The value of the renormalized $g_2(0)$ was found to be 0.46, which satisfies the generally acknowledged criterion for a single photon source ($g_2(0) < 0.5$).

## Summary

The number of experiments described and interpreted here provide a comprehensive characterisation of the optical properties of SEL in hBN. Our current understanding is that these appealing emitters originate from the optical transitions between discrete electronic states located deep in the bandgap, which emerge due to structural defects. The observation of a unique set of lines, which appear at different energies at each particular location in our samples, suggests that these defects take various forms. They may be related to boron and/or nitride vacancies, presence of additional atoms at inter-atomic sites of the lattice or combinations of both. Such defects could also induce local distortion of the lattice manifesting itself as translation or rotation of the positions of the closest atoms.

The question about the exact origin of the SEL in hBN is also connected with the puzzling observation of the emission from single defects under excitation with relatively large micrometre sized laser beam. It is counter-intuitive that among millions of illuminated atoms every so often single emitting defects are found. Our current impression, based on the statistical analysis of the distribution of the emission energy of the PL lines from multiple emitting centres and appearance of broad emission bands at some locations, is that the density of such emitting defects is highly inhomogeneous. As we have speculated, the broader structures may arise due to the optical response of the ensembles of SEL. A possible explanation of the observation of single, isolated centres may also be related, perhaps partially, to the existence of strict optical selection rules required to effectively excite the transitions within particular defects. Our preliminary study of SEL in hBN indicate that at least two conditions may apply for some transitions. One of them was discovered in the PLE measurements, where a resonant character of some PL lines was revealed. Certain lines were disappearing completely when the laser energy was tuned out of very specific, well defined values. A second property, which allows some degree of selectivity in excitation of different transitions, was uncovered in linear polarisation resolved measurements. The efficiency of the light emission from SEL in hBN was found to be highly sensitive to the direction of the linear polarisation of the laser. Each PL line exhibited its own directions of polarisation axes corresponding to the maximum and the minimum of the absorption. Therefore, combining these two features, one can imagine that in typically used excitation conditions with set wavelength and direction of linear polarisation of the laser, most of the potentially illuminated SEL would give practically none PL response.

With a large variety of presently known sources capable of SPE and operating in visible spectral range **[28-38]**, which have been studied in-depth and incorporated into heterostructures, a sound question arises if it is worth the effort and resources to bring new members into this important family of light emitters. A definite answer surely cannot be given at present time, however based on results reported so far and those brought to attention here, there is no denying that the centres in hBN giving rise to SEL constitute an appealing system. To compactly summarize their advantages, one could state that they provide robust, multicolour light sources exhibiting SPE up to room temperature, bringing together two greatly desired properties. Moreover, the utmost facility of sample preparation, requiring only deposition of hBN powder onto a supporting substrate, makes emitting centres in hBN the most accessible sources providing SPE with a great potential of building up more complex structures and devices.

## Acknowledgements

The authors acknowledge the support from the European Research Council (MOMB project No. 320590) and the EC Graphene Flagship project (No. 604391) and the ATOMOPTO project (TEAM programme of the Foundation for Polish Science co-financed by the EU within the ERDFund).

The authors of this manuscript declare no competing financial interest.

## References

[1] S. Larach, R. E. Shrader, Phys. Rev. 104 (1956) 68.

[2] A. Zunger, A. Katzir, A. Halperin, Phys. Rev. B 13 (1976) 5560.

[3] C. A. Taylor, S. W. Brown, V. Subramaniam, S. Kidner, S. C. Rand, R. Clarke, Applied Physics Letters 65 (1994) 1251.

[4] K. Watanabe, T. Taniguchi, H. Kanda, Nature Materials 3 (2004) 404.

[5] D. A. Evans, A. G. McGlynn, B. M. Towlson, M. Gunn, D. Jones, T. E. Jenkins, R. Winter, N. R. J. Poolton, J. Phys.: Condens. Matter 20 (2008) 075233.

[6] L. Museur, A. Kanaev, Journal of Applied Physics 103 (2008) 103520.

[7] A. Pierret, J. Loayza, B. Berini, A. Betz, B. Plaçais, F. Ducastelle, J. Barjon, A. Loiseau, Phys. Rev. B 89 (2014) 035414.

[8] G. Cassabois, P. Valvin, B. Gil, Nature Photonics 10 (2016) 262.

[9] R. V. Gorbachev, I. Riaz, R. R. Nair, R. Jalil, L. Britnell, B. D. Belle, E. W. Hill, K. S. Novoselov, K. Watanabe, T. Taniguchi, A. K. Geim, P. Blake, Small 7 (2011) 465.

[10] L. Museur, E. Feldbach, A. Kanaev, Phys. Rev. B 78 (2008) 155204.

[11] R. Bourrellier, S. Meuret, A. Tararan, O. Stéphan, M. Kociak, L. H. G. Tizei, A. Zobelli, Nano Lett. 16 (7) (2016) 4317.

[12] C. Palacios-Berraquero, M. Barbone, D. M. Kara, X. Chen, I. Goykhman, D. Yoon, A. K. Ott, J. Beitner, K. Watanabe, T. Taniguchi, A. C. Ferrari, M. Atatüre, Nature Communications 7 (2016) 2978.

[13] G. Clark, J. R. Schaibley, J. Ross, T. Taniguchi, K. Watanabe, J. R. Hendrickson, S. Mou, W. Yao, X. Xu, Nano Lett. 16 (6) (2016) 3944.

[14] S. Schwarz, A. Kozikov, F. Withers, J. K. Maguire, A. P. Foster, S. Dufferwiel, L. Hague, M. N. Makhonin, L. R. Wilson, A. K. Geim, K. S. Novoselov, A. I. Tartakovskii, 2D Materials 3 (2016) 025038.

[15] C. R. Dean, A. F. Young, I. Meric, C. Lee, L. Wang, S. Sorgenfrei, K. Watanabe, T. Taniguchi, P. Kim, K. L. Shepard, J. Hone, Nature Nanotechnology 5 (2010) 722.

[16] J. Xue, J. Sanchez-Yamagishi, D. Bulmash, P. Jacquod, A. Deshpande, K. Watanabe, T. Taniguchi, P. Jarillo-Herrero, B. J. LeRoy, Nature Materials 10 (2011) 282.

[17] M. Okada, T. Sawazaki, K. Watanabe, T. Taniguch, H. Hibino, H. Shinohara, R. Kitaura, ACS Nano 8 (8) (2014) 8273.

[18] L. H. Li, E. J. G. Santos, T. Xing, E. Cappelluti, R. Roldán, Y. Chen, K. Watanabe, T. Taniguchi, Nano Lett. 15 (1) (2015) 218.

[19] C. Faugeras, S. Berciaud, P. Leszczynski, Y. Henni, K. Nogajewski, M. Orlita, T. Taniguchi, K. Watanabe, C. Forsythe, P. Kim, R. Jalil, A. K. Geim, D. M. Basko, M. Potemski, Phys. Rev. Lett. 114 (2015) 126804.

[20] T. T. Tran, K. Bray, M. J. Ford, M. Toth, I. Aharonovich, Nature Nanotechnology 11 (2015) 37.

[21] T. T. Tran, C. Zacherson, A. M. Berhane, K. Bray, R. G. Sandstrom, L. H. Li, T. Taniguchi, K. Watanabe, I. Aharonovich, M. Toth, Phys. Rev. Applied 5 (2016) 034005.

[22] T. T. Tran, C. Elbadawi, D. Totonjian, C. J. Lobo, G. Grosso, H. Moon, D. R. Englund, M. J. Ford, I. Aharonovich, M. Toth, ACS Nano 10 (8) (2016) 7331.

[23] A. W. Schnell, T. T. Tran, H. Takashima, S. Takeuchi, I. Aharonovich, APL Photonics 1 (2016) 091302.

[24] N. R. Jungwirth, B. Calderon, X. Ji, M. G. Spencer, M. E. Flatté, G. D. Fuchs, Nano Lett. 16 (10) (2016) 6052.

[25] L. J. Martínez, T. Pelini, V. Waselowski, J. R. Maze, B. Gil, G. Cassabois, J. Jacques, Phys. Rev. B 94 (2016) 121405(R).

[26] N. Chejanovsky, M. Rezai, F. Paolucci, Y. Kim, T. Rendler, W. Rouabeh, F. F. de Oliveira, P. Herlinger, A. Denisenko, S. Yang, I. Gerhardt, A. Finkler, J. H. Smet, J. Wrachtrup, Nano Lett. 16 (11) (2016) 7037.

[27] A. L. Exarhos, D. Hopper, R. R. Grote, A. Alkauskas, L. C. Bassett, ACS Nano 11 (3) (2017) 3328.

[28] B. Lounis, W. E. Moerner, Nature 407 (2000) 491.

[29] A. Gruber, A. Drabenstedt, C. Tietz, L. Fleury, J. Wrachtrup, C. von Borczyskowski, Science 276 (1997) 2012.

[30] R. Brouri, A. Beveratos, J-P. Poizat, P. Grangier, Opt. Lett. 25 (2000) 1294.

[31] S. Castelletto, B. C. Johnson, V. Ivády, N. Stavrias, T. Umeda, A. Gali, T. Ohshima, Nature Materials 13 (2014) 151.

[32] M. A. Reed, J. N. Randall, R. J. Aggarwal, R. J. Matyi, T. M. Moore, A. E. Wetsel, Phys. Rev. Lett. 60 (1988) 535.

[33] P. Michler, A. Kiraz, C. Becher, W. V. Schoenfeld, P. M. Petroff, L. Zhang, E. Hu, A. Imamoglu, Science 290 (2000) 2282.

[34] M. Koperski, K. Nogajewski, A. Arora, V. Cherkez, P. Mallet, J.-Y. Veuillen, J. Marcus, P. Kossacki, M. Potemski, Nature Nanotechnology 10 (2015) 503–506.

[35] A. Srivastava, M. Sidler, A. V. Allain, D. S. Lembke, A. Kis, A. Imamoğlu, Nature Nanotechnology 10 (2015) 491–496.

[36] Y.-M. He, G. Clark, J. R. Schaibley, Y. He, M.-C. Chen, Y.-J. Wei, X. Ding, Q. Zhang, W. Yao, X. Xu, C.-Y. Lu, J.-W. Pan, Nature Nanotechnology 10 (2015) 497–502.

[37] C. Chakraborty, L. Kinnischtzke, K. M. Goodfellow, R. Beams, A. N. Vamivakas, Nature Nanotechnology 10 (2015) 507–511.

[38] P. Tonndorf, R. Schmidt, R. Schneider, J. Kern, M. Buscema, G. A. Steele, A. Castellanos-Gomez, H. S. J. van der Zant, S. Michaelis de Vasconcellos, R. Bratschitsch, Optica 2 (4) (2015) 347-352.

[39] A. V. Kuhlmann, J. Houel, A. Ludwig, L. Greuter, D. Reuter, A. D. Wieck, M. Poggio, R. J. Warburton, Nature Physics 9 (2013) 570–575.

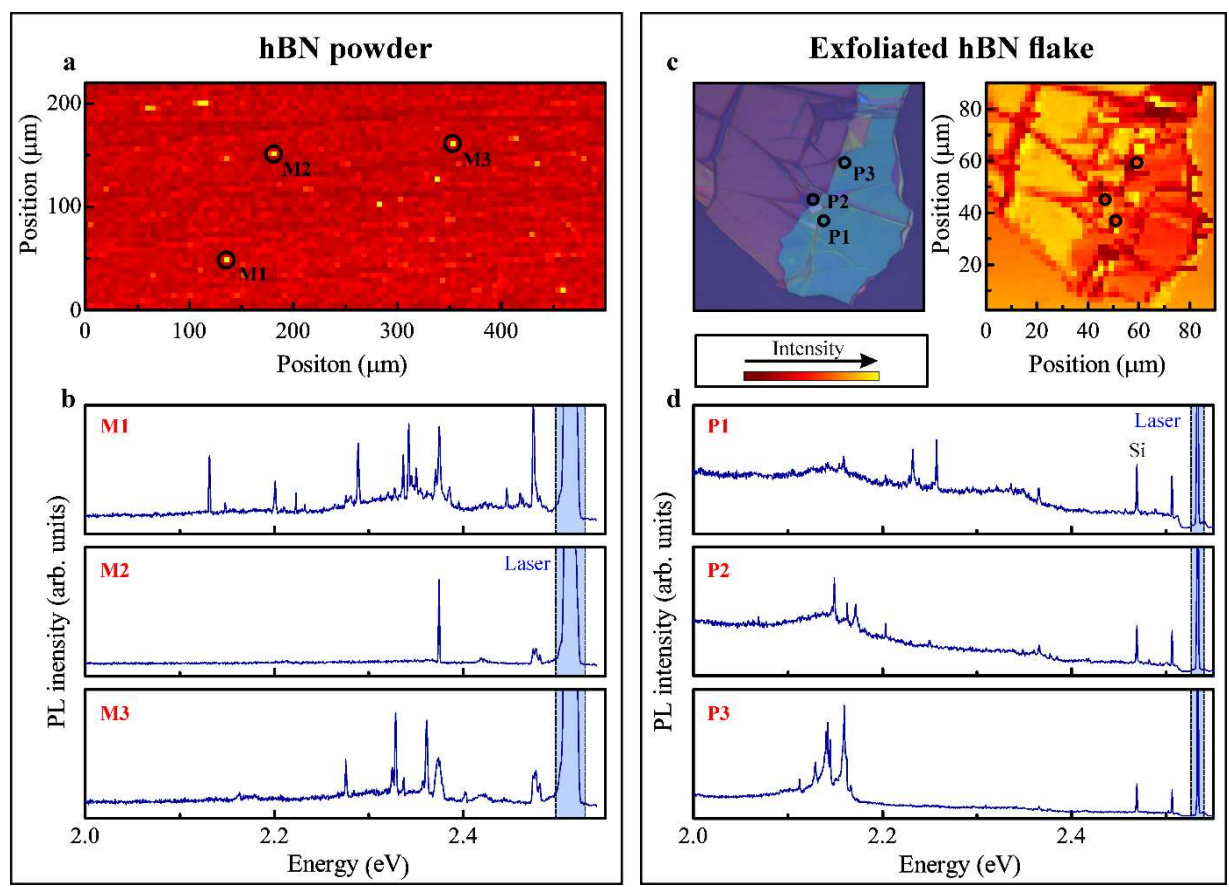

**Figure 1.** The PL mapping experiments, when the surface of the sample is scanned with a laser beam focused to a spot of about 1 μm size, are used to study systematically the appearance of emitting centres in hBN. The left panel of the figure presents results obtained for the sample covered with the hBN powder. (a) The colour map shows the intensity monitored at the specific energy of 2.374 eV, the choice of which was based on the inspection of individual spectra. The bright pixels in the map represent spots, where an emission line appears at the selected energy value. Therefore, the map does not include all the emitting centres in the area, but instead reveals particular ones. (b) Three selected spectra from spots marked in the map (M1-M3) illustrate various patterns of lines originating from the emitting centres. In the right panel, the results of mapping obtained for an exfoliated hBN flake are presented. (c) The image from the optical microscope (on the left) is confronted with a counter colour plot map of scattered laser light (on the right) to show that the details of the flake structure can be reproduced by the mapping. The emitting centres can be found in exfoliated hBN flake as well, as seen in (d) the spectra from selected spots marked in both the optical image and the map (P1-P3).

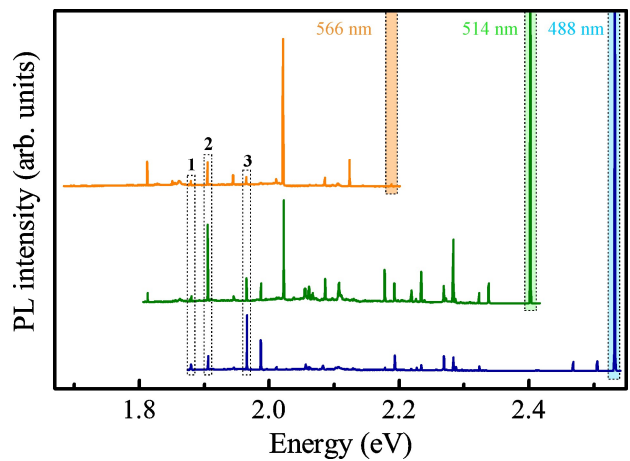

**Figure 2.** The influence of the laser energy on the photoluminescence spectra are studied for a selected emitter in hBN powder sample. The three spectra are obtained for excitation with Ar⁺ laser lines (488 nm and 514 nm) and with a rhodamine dye laser line (566 nm). Notably, the three highlighted lines exhibit different behaviour when changing the excitation energy.

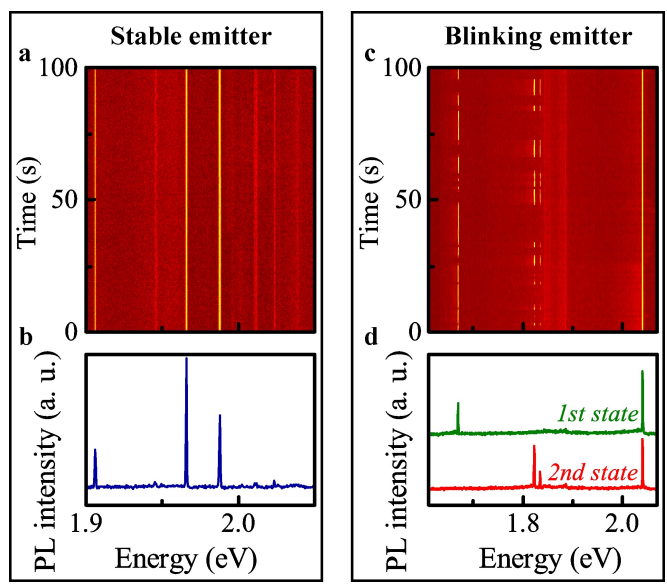

**Figure 3.** The stability of photoluminescence signal is investigated by consecutively recording the PL spectra, one after the other. (a) The temporal evolution in form of a colour map of (b) a spectrum for a selected emitting centre in hBN powder shows the existence of emitters with perfect stability regarding the intensity and emission energy. In some rare cases, the emitting centres randomly fluctuate in time between two well-defined states as seen in (c) temporal evolution of the PL spectrum of another emitter in hBN powder. Here, the highest energy line remains stable while the three lower energy lines apparently originate from two metastable states illustrated by (d) two spectra taken at different moments in time.

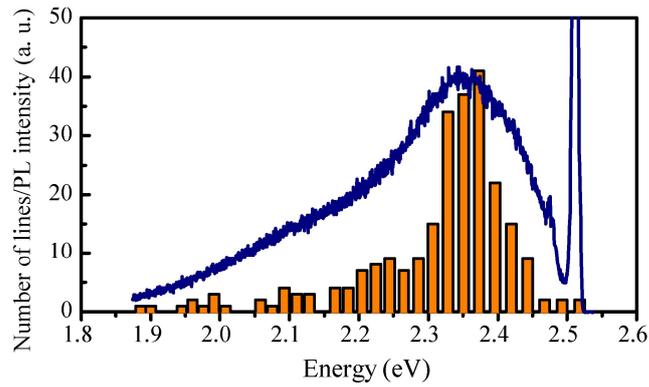

**Figure 4.** The statistical data presenting the number of emission lines at different energy values in hBN powder samples, obtained by analysing the spectra of about 80 emitters, shows a clear maximum at about 2.35 eV. The distribution of the occurrence of the narrow lines can be confronted with a spectrum showing a different kind of optical response seen in hBN powder samples. It consists of a single broad feature and appears in almost identical form at multiple places in hBN powder samples. An apparent correspondence exists between the broad band emission spectrum and the emission energy distribution of individual emitting centres.

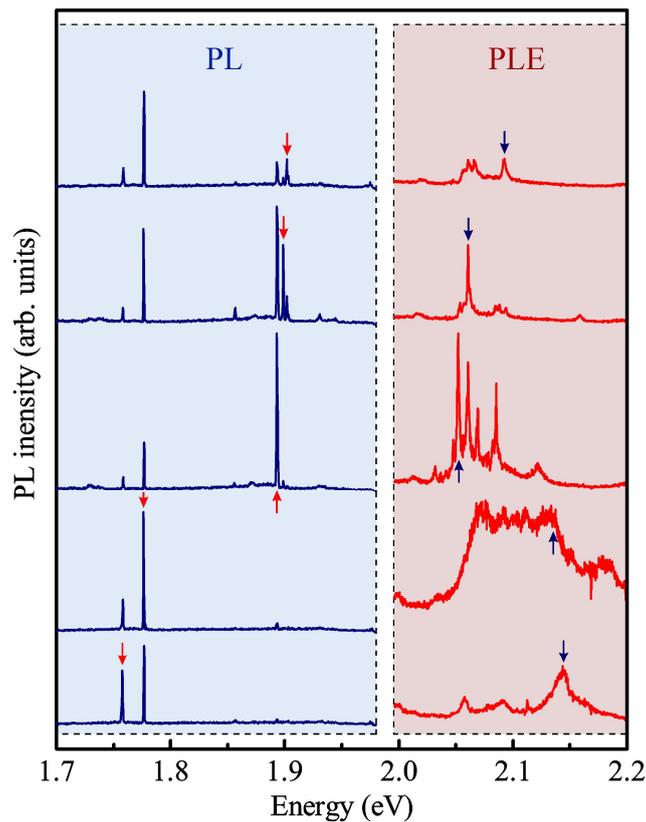

**Figure 5.** The PL spectra obtained for a selected emitting centre in hBN powder for different laser energies (blue curves) are confronted with the excitation spectra (red curves), measured with a tuneable rhodamine dye laser, to unveil a different character of the emission lines. Here, each of the excitation spectra was obtained for an individual line marked on the adjacent PL spectrum by the red arrow. Complementarily, each of the PL spectra was obtained for the laser energy marked by the blue arrow on the adjacent excitation spectrum.

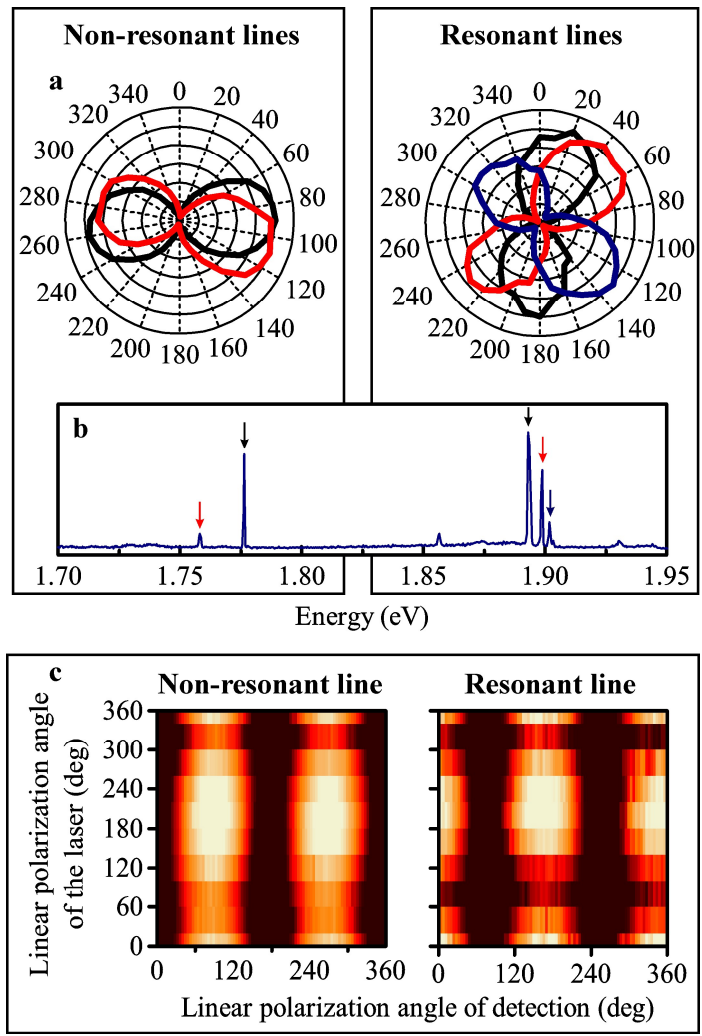

**Figure 6.** The narrow lines originating from emitting centres in hBN are usually completely linearly polarized. (a) The polar plots demonstrate the dependence of the intensity of the PL lines on the angle of the linear polarisation in detection for lines recognised as non-resonant and resonant (based on the PLE measurements). Particular curves appearing in the polar plots correspond to the lines marked with arrows in (b) the PL spectrum based on a colour code (independently for lower energy non-resonant and higher energy resonant lines). Each line has its own distinct axis of polarisation, most clearly seen when inspecting the polarisation properties of the emission signal. However, the hBN powder emitters also exhibit sensitivity to the polarisation properties of the laser used for their excitation. To illustrate that, one line from the 'non-resonant' doublet and 'resonant' triplet was selected (both lines are marked separately by black arrows in the PL spectrum) and (c) colour maps of the intensity of the lines were plotted as a function of the linear polarisation angle of the laser (vertical axis) and the detected PL signal (horizontal axis).

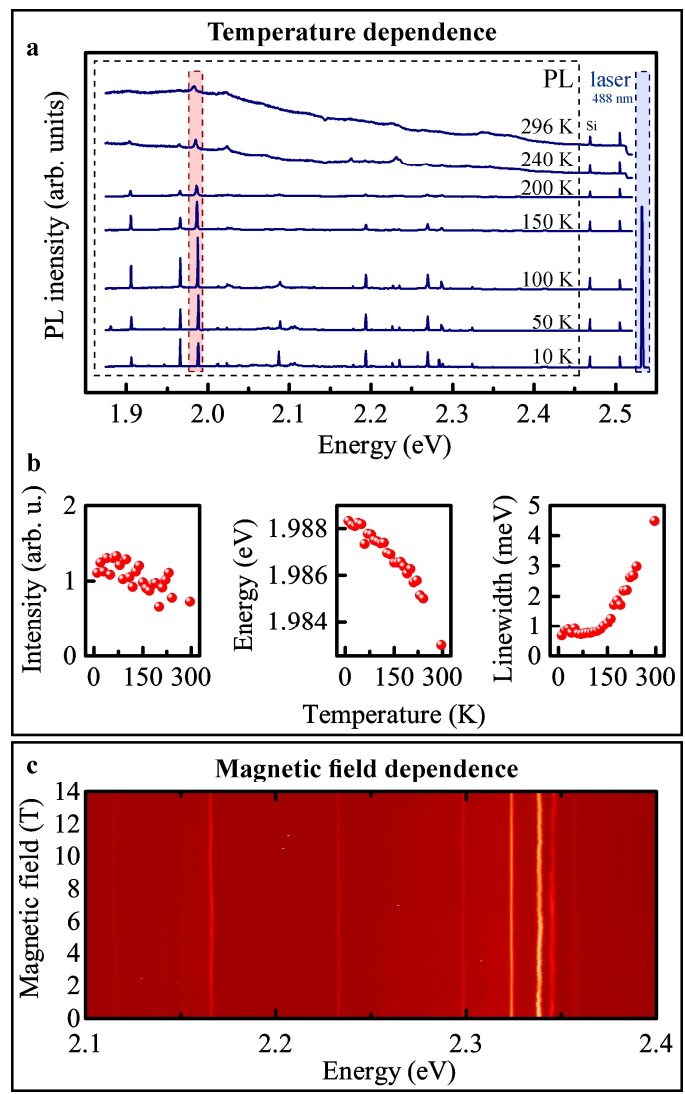

**Figure 7.** The investigation of the influence of external conditions, such as temperature or magnetic field, is a standard way to learn about the properties of light emitters in solids. Firstly, we present the PL spectra for a selected emitting centre in hBN as a function of temperature. A feature commonly appearing in defect centres in wide gap materials and present here as well is a broad band emission which gains in intensity with the increase of the temperature. Even though a presence of such spectrally broad background hinders the observation and analysis of individual lines, in some cases it is possible to trace the evolution of a particular line from helium up to room temperature, which we do here for the line at 1.988 eV (at low temperature) which is highlighted in the spectrum. For this line we present the dependence of its (b) intensity, emission energy and linewidth. (c) The measurements of the magnetic field dependence, performed on multiple randomly selected emitting spots, have proven so far that the emitting centres in hBN are completely insensitive to the application of a magnetic field in the investigated field regime (up to 14 T).

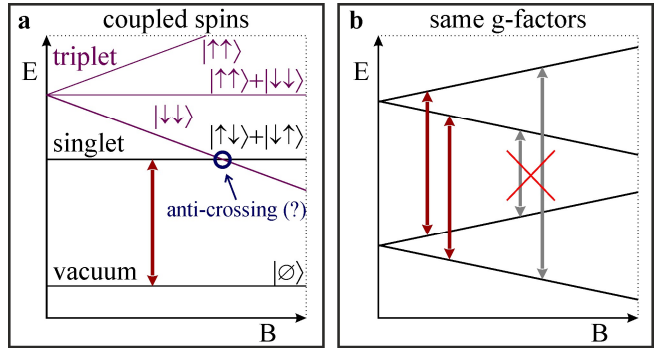

**Figure 8.** A pictorial representation of two simple cases of electronic states, which could lead to complete disappearance of the Zeeman effects for optical transitions. The first possibility is based on transition between two magnetic-field-independent states, e. g. from a vacuum state to a singlet state of two coupled electrons. This scenario is shown in the left panel of the figure, in which a presence of a higher energy triplet state is marked as well. Effects of coupling between singlet and triplet states could be revealed at higher magnetic fields. A second scenario is demonstrated in the right panel. It considers a simple case, when the same value of the Zeeman splitting of two electronic states, combined with proper selection rules, leads to field independent optical transitions.

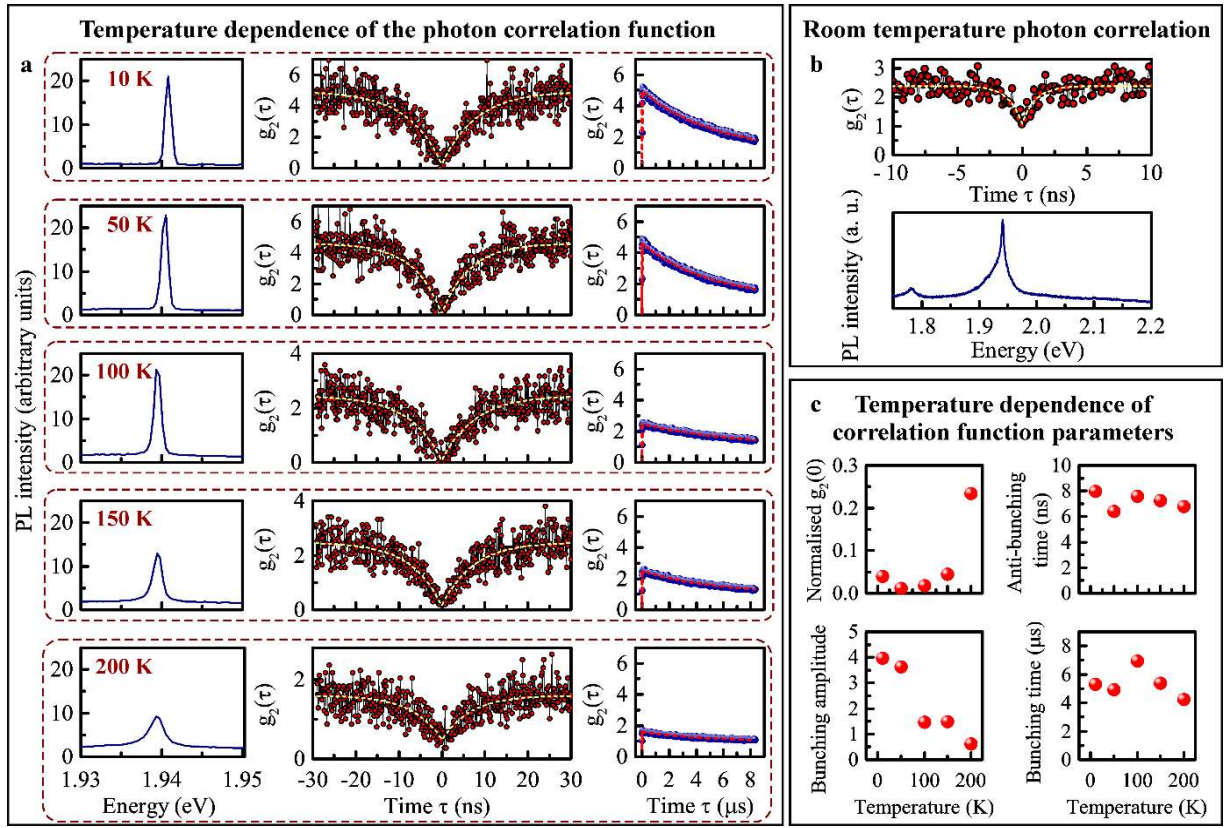

**Figure 9.** The study of photon correlations in a broad temperature range shows that SEL in hBN are remarkably good quality single photon sources. Firstly, for a single location we present (a) the temperature dependence in sub-room-temperature range (10 – 200 K) of the PL spectrum showing a selected PL line, then photon correlation function measured for this line in short timescale (tens of nanoseconds) illustrating the presence of the anti-bunching and photon correlations in long timescale (tens of microseconds) revealing an appearance of a robust bunching. For a different location (b) the photon correlation function and the PL spectrum at room temperature is shown. It is noteworthy to point out, that in order to reproduce ambient conditions of operation, the helium flow was stopped and the sample was left to thermalize to the actual temperature of the environment. Finally, (c) the basic parameters describing the short timescale anti-bunching and long timescale bunching, namely their amplitude and characteristic time, are presented as a function of temperature for the data presented in (a).